\documentstyle[12pt,epsfig]{article}
\date{}
\title{ Birefringence of high-energy $\gamma$ - quanta in the single crystals} 

\author{V.A.Maisheev \thanks{E-mail maisheev@mx.ihep.su} \\
{\it Institute for High Energy Physics, 142284, Protvino, Russia }}

\begin{document}
\maketitle
\def\arcctg{\mathop{\rm arccot}\nolimits} 
\def\ch{\mathop{\rm ch}\nolimits}
\def\sh{\mathop{\rm sh}\nolimits} 
\def\Im{\mathop{\rm Im}\nolimits}
\def\Re{\mathop{\rm Re}\nolimits}
\def\sign{\mathop{\rm sign}\nolimits}
\begin{abstract}
Problems of the experimental observation of the birefringence of
high energy $\gamma$-quanta propagating in single crystals are discussed.
\end{abstract}

\section{ Introduction}
 The birefringence of $\gamma$-quanta with energies $ >1 $ GeV propagating
in single crystals was predicted in \cite{C}. The main process by which 
$\gamma$-quanta are absorbed in single crystals is the electron-positron
pair production. The cross section of the process depends on the direction
of linear polarization of the $\gamma$-quanta relative to the 
crystallographic planes. As a result of interaction with the electric 
field of the single crystal, a monochromatic, linearly polarized beam of 
$\gamma$-quanta comprises two electromagnetic waves with different refractive
indices, so that linear polarization is transformed into circular polarization
or vice versa. This polarization phenomenon would be observed for symmetric
orientations of single crystals with respect to the direction of motion of  
$\gamma$-quanta.

The general case of the propagation of $\gamma$-quanta in single crystals
was considered in \cite{MMF,MV1,MV2}.  In these papers it was shown that
the propagating $\gamma$-beam is a superposition of the two elliptically 
polarized waves and unpolarized $\gamma$-beam obtain some degree of
circular and linear polarization after passage through a single crystal. 
In case,  
describing in \cite{C}, the beam of $\gamma$-quanta is a superposition 
of the two linearly polarized waves and  unpolarized beam obtain  
only some degree of linear polarization after propagation in single crystals.

  It is important to note that no experiments have been performed to date
to corroborate the transformation of $\gamma$-beam polarization in single
crystals, despite the notable lapse of time since the publication of 
\cite{C}. It is at least two essential purposes for experimental investigations
of the birefringence in single crystals. There are:

1)The nature of phenomenon is a manifestation of the nonlinearity of
Maxwell's equations for the electromagnetic vacuum. Of course, a single
crystal contain carriers of electric charge (electron, ions, etc), but
their direct presence is significant only if the frequencies of the 
electromagnetic radiation passing through the single crystal are low,
while  at high frequencies the fields formed by these charges play
the main role. Thus the observation of the birefringence in single 
crystals is indirect experimental proof of existence of the similar 
effect in electromagnetic vacuum (see \cite{WDHG} and the literature 
cited therein); 

2) Some possibilities exist to utilize this phenomenon  in experiments
on modern accelerators (see \cite{MV2,BT,KP} and the literature cited 
therein).

\section{ Refractive indices of $\gamma$-quanta in single crystals.}

Now we have found the refractive indices of $\gamma$-quanta propagating in 
a single crystal. Below we rewrite the components of a complex 
permittivity tensor
$\varepsilon_{ij}= \varepsilon'_{ij}+\varepsilon''_{ij} , i, \, j =1,2$ 
(the process is determined by the transverse
part of the tensor) from paper \cite{MMF}.  
Let us consider a high-energy beam of $\gamma$-quanta
moving at a small angle $\theta$ with a reciprocal lattice axis defined 
by vector $\bf{G_1}$. 
Then in the Cartesian system of coordinates such that one axis is
oriented approximately parallel to the direction of motion of the 
$\gamma$-quanta and other two axes lie in planes determined by the
vectors $\bf{G_1},\, \bf{G_2}$ and $\bf{G_1}, \, \bf{G_3}$, the   
tensor $\varepsilon_{ij}$ is a sum over reciprocal lattice vectors
${\bf{g}}= n_2 {\bf{G_2}}+n_3 {\bf{G_3}}\, (n_1=0, \theta \ll 1)$ and has
the following components:
\begin{eqnarray}
\varepsilon'_{11}={S^{'}\over 2}+ {{BN\bar{\sigma}}\over {8\pi}} 
{\hbar \over mc} \sum_{\bf{g}} \Phi(g)\,(g_2^2-g_3^2)z_g^2F'_1(z_g). 
\nonumber \\
\varepsilon'_{22}={S^{'}\over 2}- {{BN\bar{\sigma}}\over {8\pi}} 
{\hbar \over mc} \sum_{\bf{g}} \Phi(g)\,(g_2^2-g_3^2)z_g^2F'_1(z_g). 
\nonumber \\
\varepsilon'_{12}=\varepsilon'_{21}=  + {{BN\bar{\sigma}}\over {8\pi}} 
{\hbar \over mc} \sum_{\bf{g}} \Phi(g)\,(2g_2g_3)z_g^2F'_1(z_g). 
\label{1} \\
S'= 2+{{BN\bar{\sigma}}\over {\pi}}{\hbar \over mc} 
\sum_{\bf{g}} \Phi(g)\,(g_2^2+g_3^2)z_g^2F'_2(z_g,1). \nonumber \\
z_g= {{2mc^2} \over {E_{\gamma}\theta(g_2\cos\alpha +g_3 sin\alpha)}} =  
{1 \over { n_2W_V +n_3W_H}}.
\label{2}
\end{eqnarray}
The summation over {\bf{g}} satisfies the condition 
\begin{equation}
z_g>0.
\label{3}
\end{equation}
\begin{eqnarray}
\varepsilon''_{11}= {S''\over 2} - {BN \bar{\sigma} \over 16} 
{\hbar \over mc} \sum_{\bf{g}} \Phi(g)(g_2^2 - g_3^2)F''_1(z_g),
\nonumber \\
\varepsilon''_{22}= {S''\over 2} + {BN \bar{\sigma} \over 16} 
{\hbar \over mc} \sum_{\bf{g}} \Phi(g)(g_2^2 - g_3^2)F''_1(z_g),
\nonumber \\
\varepsilon''_{12}=\varepsilon''_{21}= - {BN \bar{\sigma} \over 16}
{\hbar \over mc} \sum_{\bf{g}} \Phi(g)(2g_2 g_3)F''_1(z_g),
\label{4}  \\  \nonumber
S''= \varepsilon_A+{{BN\bar{\sigma}}\over {2}}{\hbar \over mc} 
\sum_{\bf{g}} \Phi(g)\,(g_2^2+g_3^2)z_g^2F''_2(z_g,1).  
\end{eqnarray}
 The summation over {\bf{g}}  satisfies the condition
 \begin{equation}
    0<z_g \le 1
 \label{5}
 \end{equation}
The functions $F'_1, F'_2,  F''_1, F''_2, $\footnote{
Note that these functions are also used for description of the birefringence
in the laser electromagnetic wave \cite{MV1,MV}}  
 are equal to:
\begin{equation} 
{F_1}^{\prime }(z) =
\cases{ 
[\sqrt{1-z}+{z\over 2}L_-]^2+[\sqrt{1+z}-{z\over 2}L_+]^2 -
{{\pi^2z^2}\over {4}},\; 0<z \le 1,  \cr 
 -[\sqrt{z-1} -z \arcctg\sqrt{z-1}]^2+[\sqrt{1+z}- {z\over 2}L_+]^2 
, \; z>1. \cr}
\label{6}
\end{equation}
\begin{equation}
{F_2}^{\prime }(z,\mu ) =
\cases{ 
-2-\mu -(1+\mu (z-{z^2\over 2})){1\over 4} L^2_- 
-(1-\mu (z+{z^2\over 2})){1\over 4}L^2_+ + \cr
+{{(1+\mu z )\sqrt{1-z}}\over{2}}L_- -{{(\mu z-1)\sqrt{z+1}}\over {2}}L_+ 
+ {\pi^2 \over 4}(1+\mu (z-{{z^2}\over 2})),\; 0<z \le 1, \cr
-2-\mu  +(1+\mu (z-{z^2\over 2})) \arcctg^2(\sqrt{z-1}) 
-(1-\mu (z+{z^2 \over 2})){1\over 4}
L^2_+ + \cr
+ (1+\mu z)\sqrt{z-1} \arcctg\sqrt{z-1} -{{(\mu z-1)\sqrt{1+z}}\over{2}} L_+ 
,\; z>1.  \cr }
\label{7}
\end{equation}
\begin{equation}
{F_1}^{\prime \prime}(z)=
\cases{ 
z^4 (L_- + {{2\sqrt{1-z}}\over{ z}}) , \; 0< z \le 1,  \cr
0, \; z>1. \cr}
\label{8}
\end{equation}
\begin{equation}
{F_2}^{\prime \prime}(z,\mu )=
\cases{ 
z^2((1+\mu (z-{z^2\over 2}))L_- -\sqrt{1-z}(1+\mu z)), \; 0<z \le 1,  \cr
0, \; z> 1 \cr}
\label{9}
\end{equation}
 The functions   $L_+,\, L_- $ are equal to:
\begin{equation}
 L_+ =ln{{\sqrt{1+z}+1}\over {\sqrt{1+z}-1}} \, . \nonumber
\label{10}
\end{equation}
\begin{equation}
 L_- =ln{{1+\sqrt{1-z}}\over {1-\sqrt{1-z}}} \, .
\label{11}
\end{equation}
In these equations $E_\gamma$ is the energy of $\gamma$-quanta, 
$m$ is the electron mass, c is the speed of light,
$\alpha$ is
the angle between planes $({\bf{G_1}}, {\bf{G_2)}}$ 
and $({\bf{G}_1}, {\bf{K}})$,
where ${\bf{K}}$ is the momentum of $\gamma$-quanta.  
The value $\Phi(g)$ is determined by following relation:
\begin{eqnarray}
 \Phi(g)= |S({\bf{g})}|^2(1-F(g))^2\exp^{-Ag^2}/g^4 ,
\label{12}
\end{eqnarray}
where S(g) is the structure factor, F(g) is the form factor of an atom in
the single crystal and A is the mean-square amplitude of thermal vibrations
of the atoms. N is the number of atoms per unit of volume.
\begin{eqnarray}
B= {{16\pi^2} \over {N_S \Delta}}, \qquad \qquad 
\bar{\sigma}=\alpha_e Z^2 r_e^2,
 \label{13} 
\end{eqnarray}
where $ \alpha_e$ is the fine-structure constant, $r_e$ is the classical 
electron radius, $Z$ is the atomic number of the material of the single
crystal, $\Delta$ is the volume of the elementary cell and $N_S$ is the
number of atoms per this cell.
The term $\varepsilon_A$ in Eqs.(5) takes into account the absorption of
$\gamma$-quanta on the thermal vibrations of the lattice and is equal to
\begin{equation}
\varepsilon_A= {\bar{\sigma} Nc \hbar \over E_\gamma} ({2 \over 3} \psi^{am}_1
+ {1 \over 9} \psi^{am}_2),
\label{14}
\end{equation}
where the values $\psi^{am}_1$ and $\psi^{am}_2$ are approximately constants
and these quantities are determined in theory \cite{TM}.
In Eqs.(1-4,12-14) the system of units was used in which the reciprocal lattice 
constant is measured in units of $\lambda_e^{-1}$ ($\lambda_e =\hbar/mc$)
and the direct lattice constant is measured in units of $\lambda_e$;
this is adopted in the theory of coherent radiation and pair-production
\cite{TM}. 

The choice of the basic vectors ${\bf{G_1}}$, ${\bf{G_2}}$, ${\bf{G_3}}$
is not unique. It is convenient to choice these vectors along axes of
symmetry of the crystallographic lattice.  So, for instance, let choice
the vector ${\bf{G_1}}$ along the $<110>$-axis in a silicon single crystal.
Then one can choice the vectors ${\bf{G_2}}$ and ${\bf{G_3}}$ along
the $<001>$ and $<1 \bar{1} 0>$ axes, correspondingly. In this case
the lengths of these vectors are equal to
\begin{eqnarray}
 G_2=2\pi/a, \qquad \qquad G_3=2\sqrt{2}\pi/a,
\label{15}
\end{eqnarray}
where the $a$ is the side of the sell. 
 One can see from the expressions (1)-(4) that the components of the 
 tensor $\varepsilon_{ij}$ are the functions of the two universal
 parameters $W_H$ and $W_V$ (if the term $\varepsilon_A$ is ignored).  
These parameters for silicon crystal and orientation determined by Eq.(15) 
are equal to
\begin{eqnarray}
W_H=6.183 E_\gamma \theta \sin\alpha \qquad  
W_V=4.372 E_\gamma \theta \cos\alpha \qquad 
\label{16}
\end{eqnarray}
where $E_\gamma$ and $\theta$ are  measured  in  GeV  and  radians, 
correspondingly.  
 
 Knowing the permittivity tensor $\varepsilon_{ij}$ one can find
the refractive indices of $\gamma$-quanta \cite{MMF}
\begin{equation} 
\tilde n^2 
=(\varepsilon_{11}+\varepsilon_{22})/2 \pm 
\sqrt{(\varepsilon_{11}-\varepsilon_{22})^2/4 + 
\varepsilon_{12}\varepsilon_{21} }
\, ,
\label{17}
\end{equation}
Thus two waves with different indices of refraction 
$\tilde n_1$ and $\tilde n_2$  can propagate in the single crystals.
In general, these refractive indices are complex quantities.
Besides, in general case these two waves are elliptically polarized. 
However in particular case when the coordinate system exists  in which the  
tensors $\varepsilon'_{ij}$ and $\varepsilon''_{ij}$ are simultaneously
diagonal (i.e., complex tensor $\varepsilon_{ij}$ is reduced to 
principal axes) the both waves are linearly polarized. 
It is obviously that the permittivity tensor is diagonal when the
momentum of
$\gamma$-quanta lies  strictly in $({\bf{G_1}},{\bf{G_2}})$ or  
 $({\bf{G_1}},{\bf{G_3}})$ planes (angle $\alpha =0$ or $ \pi/2$).
Then the
refractive indices are equal to:
\begin{equation} 
   \tilde{n_1}= \sqrt{\varepsilon_{11}} ,\,\qquad 
   \tilde{n_2}= \sqrt{\varepsilon_{22}} 
\label{18}
\end{equation}
In this case the differences of the real and imaginary parts of
refractive indices are equal to $(\alpha=0)$ :
\begin{eqnarray}
\Re (\tilde{n_1}-\tilde{n_2})=
 {{BN\bar{\sigma}}\over {8\pi}} 
{\hbar \over mc} \sum_{\bf{g}} \Phi(g)\,(g_2^2-g_3^2)z_g^2F'_1(z_g) 
\vartheta(z_g) , \, 
\label{19}
\\
\Im(\tilde{n_1}-\tilde{n_2})=
 - {BN \bar{\sigma} \over 16} 
{\hbar \over mc} \sum_{\bf{g}} \Phi(g)(g_2^2 - g_3^2)F''_1(z_g)
\vartheta(1-z_g) \vartheta(z_g) ,
\label{20} \\
z_g=1/(n_2 W_V)
\label{21}
\end{eqnarray}
where $\vartheta$ is the Heaviside unit step function.
The similar case was considered in paper \cite{C}.

However, the $\gamma$-beam obtaining for experiments has some nonzero
phase volume and, strictly speaking,  the number of $\gamma$-quanta,
which have different angles $\theta$ but fixed angle $\alpha=0$, 
is equal to zero.
 In other words, a real $\gamma$-beam have some
distribution over the angle $\alpha$. Now we show in detail that this fact
change noticeably the relations for calculation of the refractive
indices.
 
 The components of permittivity tensor are the sum over the reciprocal
lattice vectors and summation over ${\bf{g}}$ satisfies the
conditions $z_g >0$ or $ 0<z_g<=1$ for the real and imaginary components,
correspondingly. Let us consider the first condition (for real components).
One can rewrite its in the following form: $n_2W_V +n_3W_H >0$.
When $\alpha=0$ we have $W_H=0$ and get $n_2 >0$ and $n_3$ is  an arbitrary 
integer number ($W_V \ne 0$). Now let the angle $\alpha$ is nonzero
small  angle. Then  we get
\begin{equation}
 n_2(G_2 \theta \cos{\alpha}) + n_3 (G_3 \theta \sin{\alpha}) >0  
\label{22}
\end{equation}
It easy to see that set of numbers $n_2=0, \,  n3= (1,2,3 ...) \sign{\alpha}$
($\sign$ is the function equal to $\pm 1$  according to sign of $\alpha$)
satisfies to Eq.(22).  The set of obtained numbers (in case $\alpha=0$ )
is also satisfied Eq.(22). Note that is true for any small 
nonzero angle $\alpha$.
For imaginary components the set of $n_2,\,n_3$-numbers is the same in
both cases, if only the angle $\alpha$ is enough small.

Now we can calculate the components of the permittivity tensor in 
limits $\alpha \, \rightarrow \pm \,0 $.  It is clear that permittivity
tensor (in pointed limit) have a diagonal form.
Finally we get the following    
quantity of the difference of real parts of the refractive indices: 
\begin{equation}
\Re (\tilde{n_1}-\tilde{n_2})=
 {{BN\bar{\sigma}}\over {8\pi}} 
{\hbar \over mc} \{\sum_{\bf{g}} \Phi(g)\,(g_2^2-g_3^2)z_g^2F'_1(z_g) 
\vartheta(z_g) + {8 \over 15} \sum_{n_3=1}^{\infty} \Phi(n_3 G_3)(G_3n_3)^2 \}
\label{23}
\end{equation}
Note that the left and right limits are equal in value.
The difference of the imaginary parts of refractive indices is described   
as before by Eq(20). The add term in Eq(23) is equal to the mean-square value
of the interplanar electric field (within a multiplier) \cite{MMF} .
This term is independent of the $W_H$ and $W_V$ parameters.
  
 One can pointed to the similar effect in the coherent bremsstrahlung
in single crystals. Let the electron beam motion in single crystal 
is determined by the $W_H$ and $W_V$ parameters. Then the theory
predicts that the intensity of radiation of the low energy photons is small   
enough, when $W_H=0$ and $W_V$ is reasonably large. 
However the experiments show
the significant exceeding of intensity of these photons relative to
calculated values \cite{TM,BBC}, if the calculations is not take into account 
the angular divergence of the electron beam.

\section{Influence of the $\gamma$-beam divergence on propagation }
As it was shown in  paper \cite{MMF}, in general case the $\gamma$-beam 
propagate in the single  crystal as superposition of the two elliptically
polarized waves. Birefringence  is a special case 
of propagation  of high-energy $\gamma$-quanta in single crystals,  
when the elliptical polarization of these waves (eigenfunctions of 
the problem) degenerate into  linear one. The linear polarization
point to the space symmetry of the problem as it was shown 
previously.

Now we  consider the important problem for the experimental observation
of birefringence. We want to get the answer on the following question:
Is the refractive indices and polarization states
of waves, when $\gamma$-beam move near the axis of symmetry in the 
single crystal (in other words, when $W_H \ne 0$, but $W_H \ll W_V$),
essentially changed ? 
With the aim of investigation of this problem we carry out calculations
of the refractive indices and polarization states of waves at 
the small values of $W_H$. These calculations are based on papers
\cite{MMF,MV1,MV2} where the general case of $\gamma$-quanta
propagation in the anisotropic medium was considered.
Besides, we examine only the case when the beam of $\gamma$-quanta  
move under a small angle $\theta$ with respect to one of the 
"strong" crystallographic axis
(in other words when $W_H, \, W_V \, \simeq  1)$. The difference of real 
parts of refractive indices is more significant at these orientations 
in compare with the motion of beam near crystallographic planes \cite{MMF}.

  Figures 1 and 2 show the results of calculations of the refractive
indices as functions of $W_V$ at some values of $W_H$.
One can see that the variations of the refractive indices difference
are insignificant as a whole when the parameter $W_H$ is within 0 - 0.01.
However the peaks of curve at $W_H=0$ are spreading enough when
the parameter $W_H$ rise to 0.01. The curve at $W_H=0.1$ is differ from
curve at $W_H=0$ for all practically values of the $W_V$.  
In all calculations  the Moliere form factor was employed \cite{BKF}.

Figure 3 illustrates the absolute value of circular polarization,
which have the normal electromagnetic waves (eigenfunctions of the problem).
The circular polarization $P_c$ is small when $W_H \sim 0.01 $ and
it rise to 0.5 with increasing of the parameter $W_H$ to 0.1.    
Nevertheless, the value of linear polarization $P_L=\sqrt{1-P_c^2}$ 
of the normal waves 
is dominant at all considered here values of $W_H$ and $W_V$.
Besides, the turn of the semiaxes of polarization ellipse in plane
$({\bf{G_2,G_3}})$ is take a place (see figure 4).

Thus once can say that the pure birefringence in the single crystal
takes a place for $\gamma$-beam with a small angle divergence
(the value of this divergence one can found from relation
$\delta W_H \sim 0.01$). On the other hand, the variations of polarized 
state of the $\gamma$-quanta propagating in single crystals at 
different values of $W_H$ is of more direct interest to practical goals.

 Figures 5,6 show the variation of the circular polarization of the
100 GeV $\gamma$-beam  propagating in the silicon single crystal
as a function of its thickness. The system of coordinate is chosen  so
that the Stokes parameter  of $\gamma$-beam  $\xi_3= \pm 1$  when  
100 \% linear polarization lies in planes $(1 \bar{1}0)$ and $(001)$,
correspondingly. We take for illustration the cases of partially polarized 
beam and unpolarized one at the point of entry in the single crystal. 
In the case of pure birefringence (see figure 5) the unpolarized $\gamma$-beam 
can obtain only some degree of the linear polarization
on any thickness of a single crystal (i.e. $\xi_2(x) =0$). 
In the case when
the normal electromagnetic waves is elliptically polarized the propagating
unpolarized
beam of $\gamma$-quanta can obtain some degree of the linear and circular  
polarization (see figure 6).
The transformation of linear polarization to circular one 
( under angle in $ \pm 45^{o}$ with 
respect to above-mentioned coordinate system) one can  see also on  these  
figures. The analogous curves for parameter $W_H=-0.1$ are 
mirror-symmetric with respect to x-coordinates.
The intensity of $\gamma$-beam is decreased in $\sim \, 10^9 $ 
times on  100 cm of the silicon single crystal.

Note that our consideration of the bierfrigence base on the theory
of coherent  $e^{\pm}$-pair production in single crystals \cite{TM,U}.
However this theory is violated at some orientations of single crystals
(in regions of so called "strong field") \cite{MP}. For silicon 
crystallographic planes this violation is expected at very high energy 
of $\gamma$-quanta $\gg 1 TeV$.
\section{Conclusion}
 The pure birefringence of high energy $\gamma$-beam propagating
 near crystallographic axis
(when the eigenfunctions of a problem is two linearly polarized electromagnetic
 wave) take a place
for special (predominantly symmetric) orientations. 
In general case  the propagating $\gamma$-beam is  the
superposition of two elliptically polarized waves, because of this
some peculiarities in the propagation of $\gamma$-beam exist 
even for orientations  near to the pointed symmetrical ones.
 For these orientations we can point on the following:
\newline 1. Some noticeable  degree  of  circular  polarization  of 
eigenfunctions exists.
\newline 2. Some angle shift of the axes of the polarization ellipse
takes a place also.
\newline 3. The quantities of refractive indices is changed sharply
for close orientations.
\newline 
Thus it needs to take into account these effects in experimental
observation of birefringence of a beam of $\gamma$-quanta with
some phase volume.

In addition  we consider  in detail the procedure of calculation of
refractive indices for real $\gamma$-beams.

\newpage

\begin{figure}[h]
\begin{center}
\parbox[c]{13.5cm}{\epsfig{file=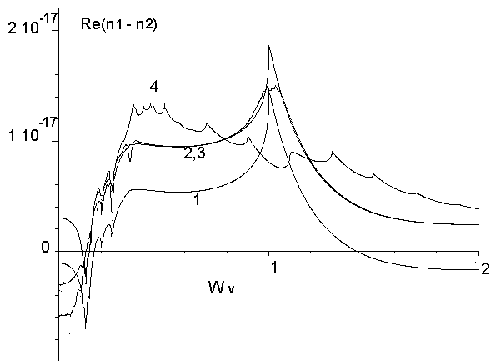,height=9cm}}
\parbox[c]{15cm}{\caption{
The differences of the real parts of the refractive indices in silicon              
as functions of $W_H$-parameter. Curves 1 and 2 are calculated at
$W_H=0$ by the use of Eq.(19) and Eq.(23), correspondingly. Curves 3 and 4 
are calculated according to Eq.(18) at $W_H=0.01$ and $0.1$. 
              }}  
\end{center} 
\end{figure}
\begin{figure}[h]
\begin{center}
\parbox[c]{13.5cm}{\epsfig{file=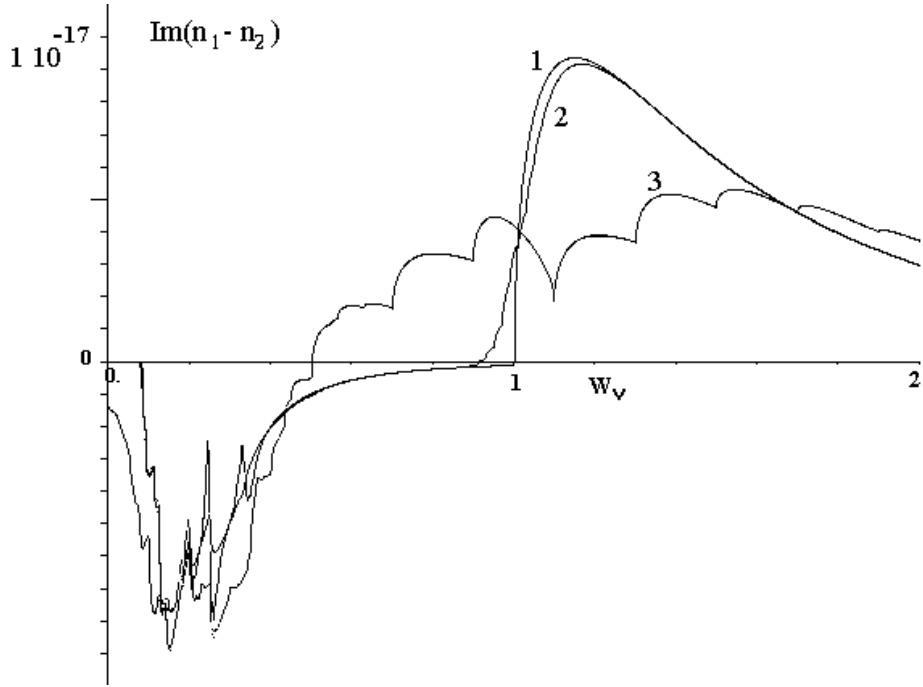,height=9cm}}
\parbox[c]{15cm}{\caption{
The differences of the imaginary parts of the refractive indices in silicon              
as functions of $W_V$-parameter. Curves 1, 2, 3 are calculated at
$W_H=0,\, 0.01, \,0.1$, correspondingly. 
              }}  
\end{center} 
\end{figure}

\begin{figure}[h]
\begin{center}
\parbox[c]{13.5cm}{\epsfig{file=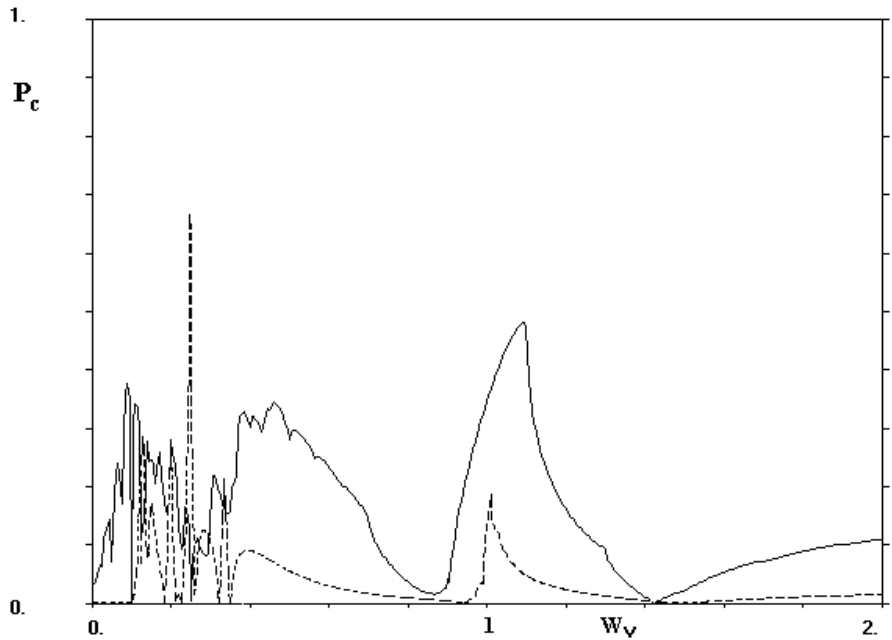,height=9cm}}
\parbox[c]{15cm}{\caption{
 Absolute value of the circular polarization  of the normal electromagnetic
waves in silicon as the function of the $W_V$-parameter. The solid curve
is calculated for $W_V= 0.1$ and the dotted line is for $W_V=0.01$.
 }}  
\end{center} 
\end{figure}

\begin{figure}[h]
\begin{center}
\parbox[c]{13.5cm}{\epsfig{file=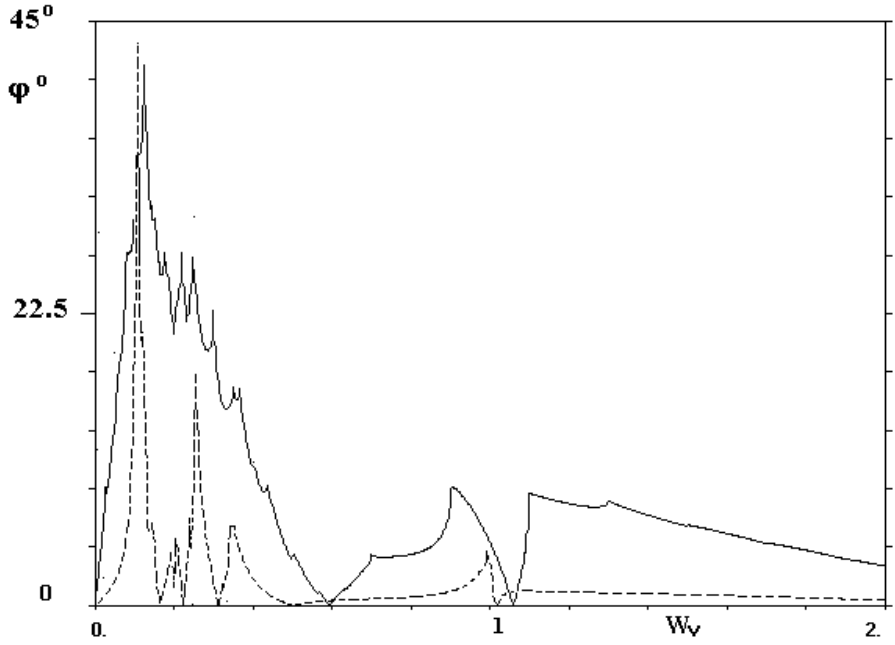,height=9cm}}
\parbox[c]{15cm}{\caption{
 Absolute value of the angle shift of the normal electromagnetic
waves in silicon as the function of the $W_V$-parameter. The solid curve
is calculated for $W_H= 0.1$ and the dotted line is for $W_H=0.01$.
 }}  
\end{center} 
\end{figure}

\begin{figure}[h]
\begin{center}
\parbox[c]{13.5cm}{\epsfig{file=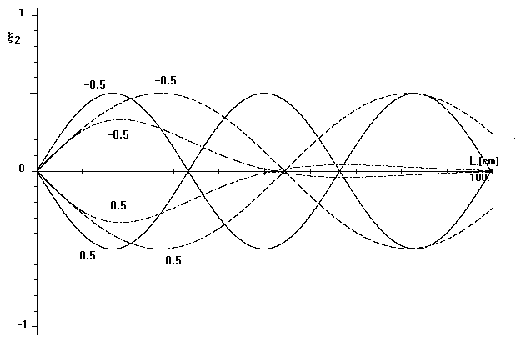,height=8.cm}}
\parbox[c]{15cm}{\caption{
The variation of circular polarization of the 100 GeV  $\gamma$-beam 
propagating in 
a silicon single crystal as function of its thickness (in centimeters).
For all curves the $W_H$-parameter is equal to 0. The parameter 
$W_V= 1., \, 0.9, \, 1.1$ for solid, dotted and dot-and-dashed curves,
correspondingly. The values near the curves are the initial quantity
of $\xi_1$ Stokes parameter for $\gamma$-beam. The other initial Stokes
parameters are equal to 0 for all curves.
 }}  
\end{center} 
\end{figure}

\begin{figure}[h]
\begin{center}
\parbox[c]{13.5cm}{\epsfig{file=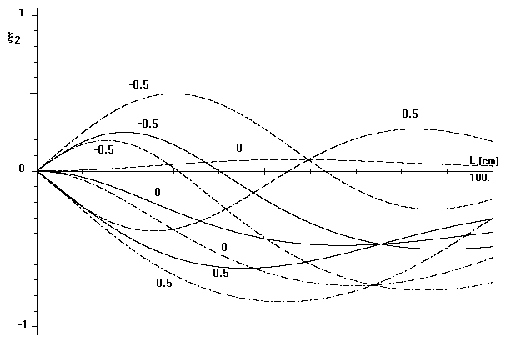,height=8.cm}}
\parbox[c]{15cm}{\caption{
The variation of circular polarization of the 100 GeV  $\gamma$-beam 
propagating in 
a silicon single crystal as function of its thickness (in centimeters).
For all curves the $W_H$-parameter is equal to 0.1. The parameter 
$W_V= 1., \, 0.9, \, 1.1$ for solid, dotted and dot-and-dashed curves,
correspondingly. The values near the curves are the initial quantity
of $\xi_1$ Stokes parameter for $\gamma$-beam. The other initial Stokes
parameters are equal to 0 for all curves.
 }}  
\end{center} 
\end{figure}

\end{document}